\documentclass[12pt]{article}
\usepackage{amsmath}
\usepackage{amssymb}
\begin{document}
\begin{center}
{\large\bf  {Pseudo-Scalar Field as a Possible Candidate for Phantom Field }}

\vskip 2.0cm

{\sf E. Harikumar$^{(a)}$, R. P. Malik$^{(b,c)}$}\\

$^{(a)}$ {\it School of Physics, University of Hyderabad,}\\
{\it Central University P. O., Gachibowli, Hyderabad - 500 046, Telangana, India}\\

\vskip 0.2cm

$^{(b)}$ {\it Physics Department, Institute of Science,}\\
{\it Banaras Hindu University, Varanasi - 221 005, U.P., India}\\

\vskip 0.2cm

$^{(c)}$ {\it DST Centre for Interdisciplinary Mathematical Sciences,}\\
{\it Institute of Science, Banaras Hindu University, Varanasi - 221 005, India}\\
{\small {\sf {e-mails:  eharikumar@uohyd.ac.in; rpmalik1995@gmail.com}}}
\end{center}

\vskip 1cm

\noindent
{\bf Abstract:}
We demonstrate the existence of a single pseudo-scalar (PS) field in the mathematically backed and parity preserving {\it modifications} of the standard
St${\ddot u}$ckelberg formalism (SSF) in the context of the Lagrangian formulation of the
(i) two (1 + 1)-dimensional (2D) massive Abelian 1-form gauge theory, (ii) three (2 + 1)-dimensional (3D) massive 
Abelian 2-form gauge theory, and (iii) four (3 + 1)-dimensional (4D) massive Abelian 3-form gauge theory. 
This PS field always turns-up with the (i) {\it negative} kinetic term, and (ii) higher order derivative terms. The {\it latter} terms are rendered into
the well-defined terms due to the imposition of the on-shell condition for the PS field
which is derived from the properly gauge-fixed Lagrangian densities for the above theories. These Lagrangian densities incorporate into themselves  the 
(i) PS field with the {\it negative} kinetic term, and (ii) pure scalar field with the {\it positive} kinetic term. However,
both these fields satisfy the Klein-Gordon (KG) equation of motion. 
The PS field (having the negative kinetic term and a well-defined mass) 
is one of the  possible candidates for the ``phantom'' field which  plays a crucial role in the cyclic, bouncing and self-accelerated 
cosmological models of the Universe. The ``exotic'' PS field {\it also} satisfies one of the criteria for being a possible candidate for dark matter.\\





\newpage

\section{Introduction}

It is an undeniable truth that there exists a deep connection between a host of ideas and concepts from pure mathematics and 
their relevance in the contexts of some of the
advanced topics in theoretical high energy physics (e.g. the modern developments in the realm of (super)string theories,
topological field theories, supersymmetric Yang-Mills theories, etc.)
 at various levels of research activities  (cf. Sec. 5, too, for more discussions). During the past few years,
we have devoted time on providing the physical realization(s) as well as meaning(s) to the abstract ideas
behind the  de Rham cohomological operators of differential geometry (see, e.g. [1-5] and references therein)
within the framework of Becchi-Rouet-Stora-Tyutin (BRST) formalism [6-9] where we have chosen (i) the simple 1D
toy models [10,11], and (ii) a set of field-theoretic models of the {\it massless} and St${\ddot u}$ckelberg-modified {\it massive} 
Abelian $p$-form ($p = 1, 2, 3$) gauge theories 
in $D = 2 p$  dimensions{\footnote{It is evident that such kinds of field-theoretic examples for Hodge theory
exist only in the {\it even} dimensions of spacetime. However, in our present endeavor, we focus on the {\it odd} dimensional 
field-theoretic model, too. For instance, we have discussed the 3D {\it modified} massive Abelian 2-form theory
(cf. Sec. 3).} ($D = 2, 4, 6 $)
 of spacetime (see, e.g. [12-14] and references therein). This kind of 
exercise (related to the models of Hodge theory) has {\it also} been performed in the contexts of some of the interesting set of 
${\mathcal N}  = 2$ SUSY quantum mechanical models with (i) a few {\it specific}  potentials of physical importance, and (ii) the
{\it general} potential (see, e.g. [15,16] and references therein).

One of the upshots of the application of BRST formalism to the above kinds of $2 p$-dimensional Abelian $p$-form
gauge theories has been the observation of a 
tower of $p$-number of fields\footnote{We have been able to
demonstrate that the (i) 2D {\it modified} massive Abelian 1-form theory, and (ii) 4D {\it modified} missive Abelian
2-form theory are endowed with a pseudo-scalar field {\it and} a pseudo-scalar field as well as an axial-vector field, 
respectively, with negative kinetic terms (but with the well defined masses). We expect that the  6D {\it modified} massive
Abelian 3-form theory will be endowed with $(i)$ a pseudo-scalar (PS) field, $(ii)$ an axial-vector field, and $(iii)$ an antisymmetric
pseudo-tensor field of rank-2 with the negative kinetic terms. Such theories  are the tractable
field-theoretic examples for Hodge theory within the framework of BRST formalism (as far as the {\it algebraic} structures are concerned).}
that possess {\it negative} kinetic terms but they satisfy the usual  Klein-Gordon (KG) equations of motion
which demonstrate that they are endowed with the well-defined {\it rest} masses  (like the real {\it massive} scalar relativistic 
particle described by the Lagrangian density for the pure scalar field with a {\it positive} kinetic term). The fields with 
negative kinetic terms (but with the well-defined masses) are a set of possible candidates for dark matter
(see, e.g. [17,18] and more references in Sec. 5) and they play a crucial role in the
contexts of the cyclic, bouncing and self-accelerated cosmological models of the Universe (see, e.g.
[19-21] and more references  in Sec. 5) where they have been called as the ``phantom" and/or ``ghost'' fields. 
From the tower of such ``exotic'' fields, it is very difficult 
to pinpoint which specific field is more fundamental (and experimentally detectable).
In our present endeavor, we try to address this issue within the purview of QFT.

The central theme of our present endeavor is to demonstrate the existence of a {\it single}
PS field with the negative kinetic term (but with a well-defined mass) in the
contexts of the Lagrangian formulation of the (i) two (1 + 1)-dimensional (2D) {\it modified} Proca theory, (ii)
three (2 +1)-dimensional (3D) {\it modified} massive Abelian 2-form theory, and
(iii) four (3 + 1)-dimensional (4D)  {\it modified} massive Abelian 3-form theory.
The PS field appears in the modifications of the standard St${\ddot u}$ckelberg formalism (SSF)
in the contexts of the {\it above} theories which are backed by the solid mathematical arguments.  We christen the modified version of the SSF
as the modified St${\ddot u}$ckelberg formalism (MSF).
One of the crucial assumptions in our present investigation is the validity of the
parity symmetry in  (i) the modifications of the SSF, and (ii) the description of {\it all} the above {\it modified} massive 
field-theoretic models (that are described by the properly gauge-fixed Lagrangian densities).

Against the backdrop of the above paragraph, it is pertinent to point out that we have been able to demonstrate 
{\it explicitly} the modification in the SSF  for the
massive Abelian $p$-form  gauge fields by invoking {\it new} compensating 
bosonic fields  (on the basis of the solid mathematical grounds).
This has been already accomplished  
in the contexts of (i) the massive 2D Abelian 1-form (i.e. Proca) theory [22], and (ii) the massive 4D Abelian 2-form theory [23].
However, the price we have paid for such kinds of modifications are (i) the {\it new} compensating fields carry the 
{\it negative} kinetic terms, and (ii)
there is appearance of the higher  (i.e. third, fourth, etc.) 
order derivative terms (besides the {\it usual} second order derivative terms) in the Lagrangian
densities  for the above {\it specific} theories where the massive gauge fields are replaced by using the MSF. In these
works [22,23], we have developed a simple 
theoretical trick to render {\it these} higher order derivative terms into well-defined terms by taking into account the 
on-shell condition (i.e. the KG equation of motion)
for the {\it new} bosonic ``exotic'' compensating field(s). The moment we observe that the new compensating bosonic field carries the negative kinetic term,
we know that it is one of the possible candidates for (i) the ``phantom'' field of the {\it specific} cosmological models of the Universe
(see, e.g. [19-21] and more references in Sec. 5), and
(ii) the dark matter (see, e.g. [17,18] more references in Sec. 5). Hence, as far as the 
relevance of these {\it new} fields in the context of the {\it latter} case is concerned, 
they must satisfy the KG equation of motion
so that they can carry the well-defined masses (which would be able to contribute to the gravitational interaction {\it alone}). This has been our logic 
behind the utilization of the KG equation to convert the {\it higher} order derivative  terms
into the well-defined terms in the specific dimensions of sapcetime  (see, e.g. [22,23] and references therein). 
We have been able to prove, ultimately, that  the (i) properly  gauge-fixed, and (ii) appropriately
well-defined Lagrangian densities (in the {\it specific} dimension of spacetime) 
produce the {\it above} on-shell condition as the Euler-Lagrange (EL) equations of motion (EoM) thereby proving the sanctity of the 
substitutions we have made to convert the higher order derivative terms into sensible and/or well-defined terms
(cf. towards the end of the {\it next} paragraph, too).

Though briefly indicated earlier, we re-emphasize that our present endeavor is {\it different} from our earlier works [22,23]
because we show the existence of a {\it single} pseudo-scalar field (with the negative kinetic term and a well-defined mass) in the
contexts of (i) the modified massive 3D Abelian 2-form gauge theory (cf. Sec. 3 below), and (ii) the modified massive 4D Abelian 3-form
gauge theory (cf. Sec. 4 below). This is {\it not} the case with our earlier works [12,23] on the modified massive 4D Abelian 2-form 
gauge theory where we have been able to show the existence of the {\it two}  ``exotic'' fields with the negative kinetic terms and well-defined
rest masses. These {\it latter} fields in [12,23] are nothing but (i) a pseudo-scalar field, and (ii) an axial-vector field. In this connection, it
is worthwhile to point out that we have been able to establish that the {\it higher} order derivative terms are {\it important} in the sense that
they lead to the precise derivations of {\it all} the terms of the appropriate 4D Lagrangian density [23] provided we use the 
on-shell conditions to convert {\it them} into sensible terms. We lay
emphasis on the fact that, in our earlier work [12], we have (i) ignored the higher derivative terms (see, e.g., Appendix in [12]), and
(ii) obtained some of the {\it appropriate} terms of the well-defined 4D Lagrangian density by the {\it trial-and-error} method.  This
is certainly {\it not} the case with our present endeavor (and our earlier works [22,23]) where we have provided the 
theoretical basis to convert the {\it higher} derivative terms into sensible terms which, ultimately, lead to the precise derivations of 
{\it all} the terms of the well-defined Lagrangian densities in the {\it specific} dimension of spacetime.

The theoretical contents of our present investigation are organized as follows. In Sec. 2, 
we recapitulate the essentials of our earlier work [22] on the modified 2D
Proca (i.e. massive Abelian 1-form gauge) theory in the gauge-fixed Lagrangian formulation
and show the existence of a {\it single} PS field with (i) the negative kinetic term, and
(ii) a well-defined rest mass. The key contents of this section 
are very {\it crucial} as they provide a very useful theoretical background for our further discussions.
Our Sec. 3 is devoted to the discussion on the 3D {\it modified} massive Abelain 2-form 
gauge theory where we show the existence of a single  PS field in the generalization of the SSF to MSF. 
Once again, the kinetic term for {\it this} field turns-up with the negative sign. In our Sec. 4,
we deal with the  4D {\it modified}
massive Abelian 3-form theory and show the existence of the {\it above}
PS field with the negative kinetic term, once again, where the SSF is generalized to MSF.
Finally, in our Sec. 5, we summarize our key results and
point out the future perspective of our present investigation.


\section{Preliminary: Modified 2D Proca Theory }

We begin with the Lagrangian density (${\cal L}_{P} $) for the D-dimensional massive Abelian 1-form (i.e. Proca)
theory, with the rest mass $m$ for the vector boson, as follows (see, e.g. [24])
\begin{eqnarray}
{\cal L}_{P} = -\; \frac{1}{4}\; F^{\mu\nu}\, F_{\mu\nu} + \frac{1}{2}\; m^2\, A^\mu\, A_\mu,
\end{eqnarray}
where $F_{\mu\nu} = \partial_\mu \, A_\nu -\partial_\nu \, A_\mu$ is the field-strength tensor and the 1-form
($A^{(1)} = A_\mu \, dx^\mu $) defines the  massive vector boson field $A_\mu$. The above Lagrangian density is endowed with
second-class constraints in the terminology of Dirac's prescription for the classification scheme of  constraints [25,26].
Hence, it does {\it not} respect any kind of gauge symmetry transformation unlike its {\it massless} version which
respects the $U(1)$ gauge symmetry. The celebrated gauge symmetry can be restored by exploiting the theoretical
potential of the standard St$\ddot u$ckelberg formalism (see, e.g. [22,24]) where the vector field $A_\mu$ is replaced by: 
\begin{eqnarray}
A_\mu \longrightarrow A_\mu \mp \frac{1}{m}\, \partial_\mu \, \phi.
\end{eqnarray}
In the above,  the field $\phi$ is a pure scalar field and the rest mass $m$ has been taken into account
in the denominator  so that the mass dimensions 
(in the natural units $\hbar = c = 1 $) of the l.h.s and r.h.s. can  match with each-other. As a consequence of 
the above replacement (2), we
obtain the St$\ddot u$ckelberg-modified Lagrangian density (${\cal L}_{S}$) as
\begin{eqnarray}
{\cal L}_{S} = -\; \frac{1}{4}\; F^{\mu\nu}\, F_{\mu\nu} + \frac{1}{2}\; m^2\, A^\mu\, A_\mu \mp \,m\, A^\mu\, \partial_\mu \phi 
+ \frac{1}{2} \partial_\mu \phi\, \partial^\mu \phi,
\end{eqnarray}
which respects (i.e. $\delta_g\, {\cal L}_{S} = 0$) the following continuous, local and infinitesimal 
gauge symmetry transformations ($\delta_g $)
\begin{eqnarray}
\delta_g \, A_\mu = \partial_\mu \Lambda, \qquad \delta_g \phi = \pm\, m \, \Lambda, \qquad \delta_g F_{\mu\nu} = 0,
\end{eqnarray}
where $\Lambda$ is the infinitesimal local gauge symmetry transformation parameter.

At this stage, a few comments are in order. First of all,
due to the introduction of the St$\ddot u$ckelberg field $\phi$, the second-class constraints of the original Proca theory
[cf. Eq. (1)] have been converted into the first-class constraints which generate the gauge symmetry transformations (4). Second.
the above gauge symmetry transformations and the invariance of the Lagrangian density 
${\cal L}_{S} $ are true, as indicated earlier, in any arbitrary D-dimension of spacetime. Third, it is straightforward to note that
the field-strength tensor $F_{\mu\nu}$ remains invariant under the replacement (2).
Finally, the kinetic term for the pure scalar field $\phi$ carries the {\it positive} sign
irrespective of the choices of the signs in the replacement (2).

In the specific case of the two (1 + 1)-dimensions of spacetime,  the replacement (2) can be {\it modified}  on the precise
mathematical grounds (see. e.g. [22] for details). The modified version of the replacement (2) due to MSF is as follows
\begin{eqnarray}
A_\mu \longrightarrow A_\mu - \frac{1}{m}\, \partial_\mu \, \phi - \frac{1}{m}\, \varepsilon_{\mu\nu}\, \partial^\nu \tilde \phi,
\end{eqnarray}
where $\tilde \phi$ is a pseudo-scaler field and $\varepsilon_{\mu\nu} $ is the antisymmetric ($\varepsilon_{\mu\nu} = - \, \varepsilon_{\nu\mu}$) 
Levi-Civita tensor in the 2D spacetime\footnote{ For the flat 2D Minkowskian spacetime, we choose the  metric tensor $\eta_{\mu\nu} =$ diag $(+1,\, -1)$
so that the dot product between two non-null vectors $P_\mu$ and $Q_\mu$ is defined as: $P \cdot Q = \eta_{\mu\nu}\, P^\mu\, Q^\nu = P_0 \, Q_0
- P_1 \, Q_1$ where the Greek indices $\mu, \nu, \lambda....= 0, 1$ stand for the time and space directions (on the above 2D flat spacetime manifold).
The 2D Levi-Civita tensor $\varepsilon_{\mu\nu}$ is chosen with the convention: $\varepsilon_{01} = +\, 1 = -\, \varepsilon^{01}$ and it satisfies
very useful  relationships: $\varepsilon^{\mu\nu}\, \varepsilon_{\mu\nu} = -\, 2!, \; 
\varepsilon^{\mu\nu}\, \varepsilon_{\mu\lambda} = -1!\; \delta^\nu_\lambda, \; 
\varepsilon^{\mu\nu}\, \varepsilon_{\nu\lambda} = +\, 1!\; \delta^\mu_\lambda$, etc.}. 
In the above equation (5), for the sake of brevity, we have chosen a {\it single} specific sign in contrast to the equation (2).
We note that the 2D field-strength tensor $F_{\mu\nu}$ possesses only a single 
independent component which is nothing but $F_{01}$ that corresponds to the electric field $E$. The {\it latter}  turns out to be a pseudo-scalar
because it is only a {\it single} object and it changes sign under the parity operation. This property can
be mathematically understood very nicely in terms of (i) the 2-form $F^{(2)}$, and (ii) the Hodge daulity $*$ operation 
on it, in the following manner
\begin{eqnarray}
F^{(2)} = d\, A^{(1)} \equiv  \frac{1}{2!}\, F_{\mu\nu}\, (dx^\mu \wedge dx^\nu), \qquad 
*\; F^{(2)} = \frac{1}{2!}\, \varepsilon^{\mu\nu}\, F_{\mu\nu} \equiv \varepsilon^{\mu\nu}\, \partial_\mu\, A_\nu,
\end{eqnarray}
where, as is evident, the 2-form $F^{(2)}$ defines the field-strength tensor $F_{\mu\nu}$ in terms of the exterior derivative 
$d = \partial_\mu \; dx^\mu$ [with $d^2 = \frac{1}{2!}\, (\partial_\mu \, \partial_\nu - \partial_\nu \, \partial_\mu) \, (dx^\mu \wedge dx^\nu) = 0$]
of the differential geometry [1-5]
and the Abelian 1-form $A^{(1)} = A_\mu \, dx^\mu $. It is straightforward to note that: 
$ *\; F^{(2)} = \varepsilon^{\mu\nu}\, \partial_\mu \, A_\nu \equiv -\, F_{01} = -\, E$ where $*$ 
is the Hodge duality operator (that is defined on the 2D flat
Minkowskian spacetime manifold). We lay emphasis on the fact that, unlike in the case of the replacement (2), we find that
the field-strength tensor $F_{01}$ does {\it not} remain invariant under the modified St$\ddot u$ckelberg-technique of replacement (5)
due to MSF. In fact, the
existing field-strength tensor $F_{01}$ transforms as follows 
\begin{eqnarray}
F_{01} \quad  \longrightarrow \quad F_{01} + \frac{1}{m}\; \big (\partial_0^2 - \partial_1^2 \big )\,\tilde \phi 
 \equiv F_{01}  + \frac{1}{m}\; \Box \,\tilde \phi,
\end{eqnarray}
where $\Box = \partial_0^2 - \partial_1^2$ is the 2D d'Alembertian operator. Thus, ultimately, we obtain the following modified form of the 
Lagrangian density (i.e. ${\cal L}^{(m)}_{S} $), namely;
\begin{eqnarray}
{\cal L}^{(m)}_{S} &=& \frac{1}{2}\, (F_{01} + \frac{1}{m}\, \Box \,\tilde \phi)^2  + \frac{1}{2}\; m^2\, A^\mu\, A_\mu - \,m\, A^\mu\, \partial_\mu \phi 
+ \frac{1}{2} \partial_\mu \phi\, \partial^\mu \phi \nonumber\\
&-& m\, \varepsilon^{\mu\nu}\, A_\mu\, \partial_\nu\, \tilde \phi 
- \frac{1}{2} \partial_\mu \tilde \phi\; \partial^\mu \tilde \phi,
\end{eqnarray}
where the superscript $(m)$ on the above Lagrangian density\footnote{It is interesting to point out that the gauge transformations (4)
are still respected by the 
action integral corresponding to the above Lagrangian density (8) under which the pseudo-scalar field does {\it not} transform at all.  In other words,
we have the continuous, local and infinitesimal gauge symmetry transformations, at this stage, as: $\delta_g \, A_\mu = \partial_\mu \Lambda, 
\; \, \delta_g \phi = +\, m \, \Lambda, \;\, \delta_g F_{\mu\nu} = 0,\;\, \delta_g \tilde \phi = 0$.}
denotes the fact that we have taken into account the modified version of 
the St$\ddot u$ckelberg-technique of replacement (5) in its derivation.
 There are a few noteworthy points at this juncture. First of all, we observe that the kinetic terms for the fields $\phi$ and $\tilde \phi$
 have opposite signs. In other words, the latter field (i.e. the PS field) has the negative kinetic term. Second, as pointed out earlier,
 the 2D field-strength tensor (i.e. $F_{01}$) does {\it not} remain invariant under the {\it modified} St$\ddot u$ckelberg-technique of replacement (5).
 Finally, we note that, for our 2D theory, we have the higher order derivative terms like: $ \sim (\Box \tilde \phi)^2$
and $ \sim F_{01}\, \Box \tilde \phi$ which contain four and three derivative terms, respectively. These terms  
 are pathological as far as the renormalizability of our theory is concerned. We can render these problematic higher order
derivative terms (see,  e.g. [22]) into the 2D sensible terms by using
the on-shell condition: $(\Box + m^2) \, \tilde \phi = 0$ so that the ``exotic'' field $\tilde \phi $ (with a {\it negative} kinetic term) 
could become like one of the possible candidates for dark matter with a well-defined rest mass  in the physical 4D spacetime (cf. Sec. 1 for more detail). 
To corroborate it, we have to show that the 
properly gauge-fixed and well-defined 
2D Lagrangian density (without any {\it higher} order derivative terms) must lead to the derivation of the {\it above} on-shell condition
  as the Euler-Lagrange (EL) equation of motion (EoM) w.r.t. the ``exotic'' PS field $\tilde \phi$ (with the negative kinetic term).

Taking into account the definition $F_{01} = E$ and the on-shell condition: $\Box \, \tilde \phi = -\, m^2\, \tilde \phi $, 
we obtain the 2D well-defined
Lagrangian density in the 't Hooft gauge (see, e.g. [24])  as 
\begin{eqnarray}
{\cal L}^{(m)}_{S}  + {\cal L}_{gf} &=& \frac{1}{2}\, (E - \, m\, \,\tilde \phi)^2  
+ \frac{1}{2}\; m^2\, A^\mu\, A_\mu - \,m\, A^\mu\, \partial_\mu \phi 
+ \frac{1}{2} \partial_\mu \phi\, \partial^\mu \phi \nonumber\\
&-& m\, \varepsilon^{\mu\nu}\, A_\mu\, \partial_\nu\, \tilde \phi 
- \frac{1}{2} \partial_\mu \tilde \phi\, \partial^\mu \tilde \phi - \frac{1}{2} \, \bigl [(\partial \cdot A) + m\, \phi \bigr]^2, 
\end{eqnarray}
where ${\cal L}_{gf}$ is the gauge-fixing term. The above total Lagrangian density  
 does {\it not} contain any {\it higher} order derivative terms. The EL-EoMs from
 the above {\it properly} gauge-fixed  Lagrangian density, for the pure scalar field $\phi$ and 
pseudo-scalar $\tilde \phi$, are as follows:
\begin{eqnarray}
\partial_\mu \;  \Bigl [ \frac{\partial\, \big ({\cal L}^{(m)}_{S}  + {\cal L}_{gf} \big )}{\partial (\partial_\mu \phi)} \Bigr ] 
=  \frac{\partial\, \big ({\cal L}^{(m)}_{S} + {\cal L}_{gf} \big)}{\partial \phi} &\Longrightarrow& 
\Box \, \phi - m\, (\partial \cdot A) = - m\, (\partial \cdot A) - m^2 \, \phi,\nonumber\\
\partial_\mu \; \Bigl [ \frac{\partial\, \big ({\cal L}^{(m)}_{S}  + {\cal L}_{gf} \big )}{\partial (\partial_\mu \tilde \phi)} \Bigr ]
=  \frac{\partial\, \big ({\cal L}^{(m)}_{S} + {\cal L}_{gf} \big)}{\partial \tilde \phi} &\Longrightarrow& 
- \Box \, \tilde \phi + m\, \varepsilon^{\mu\nu}\, \partial_\mu\, A_\nu =   - m\, E + m^2 \, \tilde \phi.
\end{eqnarray}
A close look at the above EL-EoMs establishes that we have already derived the on-shell conditions for the 
pure scalar field $\phi$
and PS field $\tilde \phi$ as: $(\Box + m^2) \, \phi = 0 $ and $(\Box + m^2) \, \tilde \phi = 0$, 
respectively. In the derivation
of the {\it latter}, we have used the definition of the 2D electric  field $E = F_{01} = -\, \varepsilon^{\mu\nu}\, \partial_\mu\, A_\nu$.
Thus, it is clear that our {\it physically} motivated choice of the on-shell condition 
(i.e. $\Box \, \tilde \phi = -\, m^2\, \tilde \phi $) for the PS field,  in the derivation of the 2D
well-defined Lagrangian  density (9) from (8), is {\it correct}.

We end this section with the remark that 
the gauge-fixed Lagrangian density (9) has been generalized to the quantum level
within the framework of Becchi-Rouet-Stora-Tyutin (BRST) 
formalism in our earlier work (see, e.g. [27,28,22] and references therein) and  it has been established that the 
2D modified Proca theory is a tractable field-theoretic model for the Hodge theory
where the symmetries and conserved charges provide the physical realizations of
the de Rham cohomological operators of differential geometry [1-5] at the {\it algebraic} level. Second, we have been able to show the existence of
a couple of useful discrete duality transformations and {\it six} continuous transformations for 
{\it this} 2D theory within the framework
of BRST formalism. Thus, if the beauty of a theory is determined and defined in terms of 
the numbers and varieties of symmetries it respects, {\it this} 2D model belongs to this beautiful category. 
Third, the {\it higher} derivative terms of Eq. (8) are very {\it important} because the properly well-defined Lagrangian density
(9) has been derived from (8) by using the on-shell condition for the ``exotic'' PS field. Exactly the same 
Lagrangian density [i.e. Eq. (9)] has been derived in our our earlier works [27,28] by the
trial-and-error method which establishes (i) the {\it utility} and importance of the higher order 
derivative terms in (8), and (ii) the usefulness of the simple
theoretical trick we have adopted to render the {\it higher} 
order derivative terms into the 2D sensible and/or well-defined terms. Finally, the existence of the
negative kinetic term for the {\it massive} PS field, on the basis of symmetry grounds {\it alone}, 
establishes it as a possible candidate for dark matter (like the real dark matter of 4D spacetime (see, e.g. [17,18] for details)). 
Such kinds of ``exotic'' fields have also become 
quite popular in the realm of the cosmological models of the Universe.
In particular, the bouncing, cyclic and self-accelerated cosmological
models of the Universe crucially depend on the fields with negative kinetic terms (see, e.g. [19-21]). \\ 



\section{Modified Massive 3D Abelian 2-Form Theory}

We begin with the D-dimensional Kalb-Ramond Lagrangian density 
${\cal L}_{(0)}$ for the free Abelian 2-form  {\it massive} theory, with the rest mass equal to $m$, as follows (see, e.g. [12,23])
\begin{eqnarray}
{\cal L}_{(0)} = \frac{1}{12}\, H^{\mu\nu\lambda} H_{\mu\nu\lambda} - \frac{m^2}{4}\, B^{\mu\nu} B_{\mu\nu},
\end{eqnarray}
where the antisymmetric $(B_{\mu\nu} = - B_{\nu\mu})$ tensor field $B_{\mu\nu}$ is the massive Abelian 2-form 
$[B^{(2)} = \frac{1}{2!}\, B_{\mu \nu}\, (dx^\mu \wedge dx^\nu)]$ field and the curvature (i.e. field-strength) tensor 
$H_{\mu\nu\lambda} = \partial_\mu B_{\nu\lambda} + \partial_\nu B_{\lambda\mu} + \partial_\lambda B_{\mu\nu}$ is derived  
from the 3-form $ [H^{(3)} = d B^{(2)} \equiv \frac{1}{3!}\,\, H_{\mu\nu\lambda} \, (dx^\mu \wedge dx^\nu \wedge dx^\lambda)]$
 where $d  = \partial_\mu\; d x^\mu$ (with $d^2= 0$) is the exterior derivative of differential geometry. This Lagrangian density 
is endowed with the second-class constraints and, therefore, there is {\it no} gauge symmetry transformation in the theory.
However, we can restore the gauge symmetry transformation by exploiting the theoretical potential of the  St$\ddot u$ckelberg-technique of 
replacement where the antisymmetric tensor $B_{\mu\nu} $ field is redefined as
\begin{eqnarray}
B_{\mu\nu}  \longrightarrow  B_{\mu\nu} - \frac{1}{m}\, \big(\partial_\mu \phi_\nu - \partial_\nu \phi_\mu) 
\equiv  B_{\mu\nu} - \frac{1}{m}\, \Sigma_{\mu\nu},
\end{eqnarray}
where the 2-form $\Sigma^{(2)} = \frac{1}{2!}\, \Sigma_{\mu\nu}\, (dx^\mu \wedge dx^\nu)$ defines the field strength tensor
$\Sigma_{\mu\nu} = (\partial_\mu \phi_\nu - \partial_\nu \phi_\mu) $ for the 1-form $\Phi^{(1)} = \phi_\mu\, dx^\mu$ 
Lorentz vector field $\phi_\mu$. The substitution of  (12) into (11) leads to the following form the St$\ddot u$ckelberg-modified
Lagrangian density (${\cal L}_{(0)}^{(S)}$):
\begin{eqnarray}
{\cal L}_{(0)}^{(S)}  = \frac{1}{12}\, H^{\mu\nu\lambda} H_{\mu\nu\lambda} - \frac{m^2}{4}\, B^{\mu\nu} B_{\mu\nu} 
- \frac{1}{4}\, \Sigma^{\mu\nu} \Sigma_{\mu\nu}  
+ \frac{m}{2}\, B^{\mu\nu} \Sigma_{\mu\nu}. 
\end{eqnarray}
The above St$\ddot{u}$ckelberg-modified Lagrangian density respects (i.e. $\delta_g {\cal L}_{(0)}^{(S)} = 0$) 
the following continuous, local  
and infinitesimal gauge symmetry transformations $(\delta_g)$
\begin{eqnarray} 
\delta_g B_{\mu\nu}  = - \,(\partial_\mu  \Lambda_\nu - \partial_\nu \Lambda_\mu), 
\quad \delta_g \phi_\mu = (\partial_\mu\,\Lambda -\, m\, \Lambda_\mu), \quad 
\delta_g  \Sigma_{\mu\nu} = -\, m\, (\partial_\mu  \Lambda_\nu - \partial_\nu \Lambda_\mu),
\end{eqnarray}
where $\Lambda_\mu$ and $\Lambda$ are the 
infinitesimal Lorentz vector and scalar gauge transformation parameters, respectively,
and it is obvious that  the field-strength tensor is a gauge invariant (i.e. $\delta_g\,H_{\mu\nu\lambda} = 0 $)
quantity. It goes without saying that 
(i) the introduction of the St$\ddot{u}$ckelberg field $\phi_\mu$ has converted the second-class constraints of the original Lagrangian density (11) 
into the first-class variery in the terminology of the Dirac prescription for the classification scheme of constraints [25,26], and
(ii) the field-strength tensor $H_{\mu\nu\lambda}$ remains invariant (i.e. $H_{\mu\nu\lambda} \rightarrow H_{\mu\nu\lambda} $) under
the St$\ddot{u}$ckelberg technique of replacement (12). This observation (which is valid
for any arbitrary D-dimensional modified massive Abelian 2-form theory) is very crucial for our further discussions in the context of
3D modified massive Abelian 2-form theory where the pseudo-scalar field appears in the {\it modified} version of (12).

We concentrate now on the  three (2 + 1)-dimensional (3D) massive Abelian 2-form theory\footnote{We choose the flat 3D metric
tensor $\eta_{\mu\nu} =$ diag $ (+1, -1, -1)$ so that the dot product between two non-null vectors $S_\mu$ and $T_\mu$ is
defined as: $S \cdot T = \eta_{\mu\nu} \, S^\mu\, T^\nu = S_0\, T_0 - S_i\, T_i$ where the Greek indices $\mu, \nu, \lambda...= 0, 1, 2$ 
stand for the time and space directions on the 3D Minkowskian spacetime manifold and the Latin indices $i, j, k...= 1, 2$ 
correspond to the space directions only. The Einstein summation convention is adopted everywhere in our entire text. The convention for the 
3D Levi-Civita tensor is chosen to be such that: $\varepsilon_{012} = + \,1 =  \varepsilon^{012}$ and the other useful
relationships are: $\varepsilon^{\mu\nu\lambda}\, \varepsilon_{\mu\nu\lambda}
= 3!, \;\; \varepsilon^{\mu\nu\lambda}\, \varepsilon_{\mu\nu\rho} = 2!\; \delta^\lambda_\rho$, etc.}
 where the {\it standard} St$\ddot{u}$ckelberg-technique
of replacement (12) is generalized in the following fashion to its {\it modified} version (due to MSF), namely;
\begin{eqnarray}
B_{\mu\nu}  \longrightarrow   B_{\mu\nu} - \frac{1}{m}\, \Sigma_{\mu\nu} - \frac{1}{m}\, \varepsilon_{\mu\nu\lambda}\, \partial^\lambda\, \tilde \phi,
\end{eqnarray}
where $\tilde \phi $ is the pseudo-scalar field and $\varepsilon_{\mu\nu\lambda}$ is the totally antisymmetric 3D Levi-Civita tensor where
we have assumed that there is {\it no} parity violation in our theory. As a consequence of (15), it turns out that the field-strength tensor
acquires an additional term. To comprehend it clearly, we have to note the following peculiarities of the massive 3D Abelian 2-form theory.
First, we observe that the field-strength tensor ($H_{\mu\nu\lambda}$) has only a {\it single}
	independent component $H_{012}$. As a result, the
kinetic term for the gauge field is:
\begin{eqnarray}
\frac{1}{12}\, H^{\mu\nu\lambda}\, H_{\mu\nu\lambda} = \frac{1}{2}\, H^{012} \,H_{012} \equiv \frac{1}{2}\, H_{012} \,H_{012}.
\end{eqnarray}
Second, from the definition of the 3-form $H^{(3)} = d\, B^{(2)}$, it is clear that we have the following
\begin{eqnarray}
H^{(3)} = d B^{(2)} \equiv \frac{1}{3!}\,\, H_{\mu\nu\lambda} \, (dx^\mu \wedge dx^\nu \wedge dx^\lambda), \qquad
*\; H^{(3)} = \frac{1}{3!}\, \varepsilon^{\mu\nu\lambda} \, H_{\mu\nu\lambda} \equiv H_{012},
\end{eqnarray}
where $*$ is nothing but the Hodge duality operator [1-5]
on the 3D flat Minkowskian spacetime manifold. The {\it last} entry in the above equation demonstrates
that the component $H_{012}$ is  a pseudo-scalar field (because of the presence of the Levi-Civita tensor). Finally, we note that the  above
component $H_{012}$ transforms, due to the replacement (15), as:
\begin{eqnarray}
H_{012} \longrightarrow H_{012} - \frac{1}{m}\; (\partial_0^2 - \partial_1^2 - \partial_2^2)\, \tilde \phi \;\equiv\; 
 H_{012} - \frac{1}{m}\; \Box\, \tilde \phi.
\end{eqnarray}
Hence the kinetic term in (16) for the gauge field (in the case  of our 3D 
{\it modified} massive Abelian 2-form theory) transforms as follows
\begin{eqnarray}
\frac{1}{2}\, H_{012} \,H_{012} \longrightarrow \frac{1}{2}\, (H_{012} - \frac{1}{m}\; \Box\, \tilde \phi)^2,
\end{eqnarray}
which demonstrates that we have
the higher order derivative terms in our theory (due to the above expression for the kinetic term)
where, apart from the {\it usual} two derivative terms, there are  
three and four derivative terms, too. We can render these higher derivative terms into 3D 
sensible terms by exploiting the on-shell 
condition ($\Box\, \tilde \phi = -\, m^2\, \tilde \phi $) for the pseudo-scalar field (as discussed in Sec. 2). However,
we shall demonstrate that the well-defined 3D Lagrangian density
(with proper gauge-fixing terms) will produce {\it this} on-shell condition as the
EL-EoM w.r.t. the pseudo-scalar field $\tilde \phi$ of our theory [cf. Eq. (22) below].

Against the backdrop of the above discussions, we obtain the following {\it modified} version of the Lagrangian density (${\cal L}_{(0)}^{(MS)}$)
which is well-defined\footnote{It is worthwhile to point out that the action integral, corresponding to the Lagrangian density 
${\cal L}_{(0)}^{(MS)}$, respects the continuous, local and infinitesimal 
gauge symmetry transformations: $\delta_g B_{\mu\nu}  = - \,(\partial_\mu  \Lambda_\nu - \partial_\nu \Lambda_\mu), 
\; \delta_g \phi_\mu = (\partial_\mu\,\Lambda -\, m\, \Lambda_\mu), \; 
\delta_g  \Sigma_{\mu\nu} = -\, m\, (\partial_\mu  \Lambda_\nu - \partial_\nu \Lambda_\mu), \; 
\; \delta_g H_{\mu\nu\lambda} = 0, \; \delta_g \tilde \phi = 0 $ where
the pseudo-scalar field $\tilde \phi $ remains unaffected by the gauge symmetry transformations.} (without any higher order derivative terms) on the 
3D Minkowskian flat spacetime manifold, namely;
\begin{eqnarray}
{\cal L}_{(0)}^{(MS)}  &=& \frac{1}{2}\, (H_{012} + m\, \tilde \phi)^2 - \frac{m^2}{4}\, B^{\mu\nu} B_{\mu\nu} 
+ \frac{m}{2}\, B^{\mu\nu} \big [\Sigma_{\mu\nu} + \varepsilon_{\mu\nu\lambda}\, \partial^\lambda \, \tilde \phi \big ] \nonumber\\
&-&\, \frac{1}{2}\, \partial_\mu \tilde \phi\, \partial^\mu \tilde \phi - \frac{1}{4}\, \Sigma^{\mu\nu} \Sigma_{\mu\nu},  
\end{eqnarray}
modulo  a total spacetime derivative: $\partial_\mu  \, [- \, \varepsilon^{\mu\nu\lambda}\, \phi_\nu \, \partial_\lambda \, \tilde \phi ]$ which does 
not play any significant role in the description of the dynamics of our theory. The superscript $(MS)$ on the above Lagrangian density
denotes that we have exploited the beauty and strength of the replacement (15) which is nothing but the {\it modified} version of the
St$\ddot{u}$ckelberg-technique of the replacement (12). We can incorporate, at this juncture, the appropriate gauge-fixing terms for the 
gauge field $B_{\mu\nu}$ as well as the St$\ddot{u}$ckelberg vector field $\phi_\mu$ as follows
\begin{eqnarray}
{\cal L}_{(0)}^{(MS)} + {\cal L}_{gf}^{(B)} &=& \frac{1}{2}\, (H_{012} + m\, \tilde \phi)^2 - \frac{m^2}{4}\, B^{\mu\nu} B_{\mu\nu} 
+ \frac{m}{2}\, B^{\mu\nu} \big [\Sigma_{\mu\nu} + \varepsilon_{\mu\nu\lambda}\, \partial^\lambda \, \tilde \phi \big ] 
- \frac{1}{4}\, \Sigma^{\mu\nu} \Sigma_{\mu\nu} \nonumber\\
&-&\, \frac{1}{2}\, \partial_\mu \tilde \phi\, \partial^\mu \tilde \phi 
- \frac{1}{2} \, (\partial \cdot \phi + m\, \phi)^2 + \frac{1}{2}\, \big(\partial^\nu B_{\nu\mu} - \partial_\mu\, \phi + m \, \phi_\mu \bigr )^2,
\end{eqnarray}
where the last {\it two} terms are the gauge-fixing terms where the mass dimensions (in the natural units) 
of all the relevant fields have been taken into account (see, e.g. [12-14] for details). The superscript $(B)$ on the gauge-fixing term
denotes that {\it this} term is for the massive Abelian 2-form gauge field $B_{\mu\nu}$.
We  are now in the position to obtain the EL-EoMs 
from the above properly gauge-fixed Lagrangian  density (21), namely; 
\begin{eqnarray}
 -\, \partial_\mu \big[\partial_\nu \, B^{\nu\mu} - \partial^\mu \phi + m \, \phi^\mu \big ] = - m\, (\partial \cdot \phi) + m^2 \, \phi
&\;\;\Rightarrow \;\;& \big (\Box + m^2 \big)\, \phi = 0, \nonumber\\
 \partial_\mu \big[-\, \partial^\mu \, \tilde \phi + \frac{m}{2}\, \varepsilon^{\mu\nu\lambda}\, B_{\nu\lambda} \big]
= +\,  m \, \big (H_{012} + m \, \tilde \phi \big) &\;\;\Rightarrow \;\;& \big (\Box + m^2 \big)\, \tilde \phi = 0,
\end{eqnarray}
where we have taken into account the inputs from (17). To be precise, we note that we have already shown that:
$H_{012} = \frac{1}{3!}\, \varepsilon^{\mu\nu\lambda}\, H_{\mu\nu\lambda} \equiv \frac{1}{2!}\, \varepsilon^{\mu\nu\lambda}\, 
(\partial_\mu\, B_{\nu\lambda})$. Thus, it is very interesting to point out that the scalar and the ``exotic'' PS fields 
obey the KG equation of motion even though their kinetic terms have {\it opposite} signs in our theory. The
KG equation of motion encodes in itself the description of a  massive {\it real} spin-0 scalar relativistic particle. Thus, it is 
very exciting to note that the ``exotic'' PS field possesses a well-defined rest mass (just like 
a {\it real}  massive relativistic  spin-0 scalar particle) even though it is endowed 
with the {\it negative} kinetic term in our theory.

We wrap-up this section with the following remarks. First of all, the kinetic term for the pure scalar field
carries a positive sign and it comes out elegantly from the gauge-fixing term for the massive Abelian 2-form  gauge field. On the
contrary, we have already noted that the kinetic term for the ``exotic'' PS field is endowed with the negative sign. 
However,
{\it both} the fields obey the KG equation of motion which is the signature of a massive 
{\it real} relativistic particle (of rest mass $m$).
Hence, the PS field of our theory is a possible candidate like dark matter
of physical 4D spacetime  (see, e.g.  [17,18] and references in Sec. 5). Second, the above
properly gauge-fixed Lagrangian density, for the modified 3D massive Abelian 2-form theory,
can be generalized to its quantum version (within the framework of BRST formalism)
as we have done for the modified 2D Proca theory [27,28,22] and its various symmetries can be studied. 
 Finally, the PS field corresponds to  an ``exotic'' particle in the sense that it possesses negative kinetic term (but 
endowed with a well-defined  mass). \\

\vskip 1.5cm


\section{Modified Massive 4D Abelian 3-Form Theory}

With the background of the {\it last} two sections, we would like to be brief in the description of our present section
where the D-dimensional Lagrangian density [${\cal L}_{(A)} $] for the massive Abelian 3-form theory (with the rest mass equal to $m$)
is studied. The appropriate Lagrangian density for this theory is (see, e.g.  [29] for details)
\begin{eqnarray}
{\cal L}_{(A)} =-\, \frac{1}{48}\, H^{\mu\nu\lambda\zeta}\, H_{\mu\nu\lambda\zeta} + \frac{m^2}{12}\, A^{\mu\nu\lambda}\, A_{\mu\nu\lambda},
\end{eqnarray}
where the Abelian  3-form $[A^{(3)} = \frac{1}{3!}\, A_{\mu \nu\lambda}\, (dx^\mu \wedge dx^\nu \wedge dx^\lambda)]$ defines
the totally antisymmetric Abelian massive field $A_{\mu \nu\lambda} $ and the 4-form 
$ [H^{(4)} = d\, A^{(3)} \equiv \frac{1}{4!}\,\, H_{\mu\nu\lambda\zeta} \, (dx^\mu \wedge dx^\nu \wedge dx^\lambda \wedge dx^\zeta)]$ leads to 
the derivation of the field-strength tensor:
$H_{\mu\nu\lambda\zeta} = \partial_\mu A_{\nu\lambda\zeta} - \partial_\nu A_{\lambda\zeta\mu} 
+ \partial_\lambda A_{\zeta\mu\nu} - \partial_\zeta A_{\mu\nu\lambda}$
where $d$ is the exterior derivative of differential geometry [1-5]. The above Lagrangian density does {\it not} respect any 
gauge symmetry transformation because it is endowed with the second-class constraints. However, the following St$\ddot u$ckelberg-technique of 
replacement
for the totally antisymmetric massive field $A_{\mu\nu\lambda}$, namely;
\begin{eqnarray}
A_{\mu\nu\lambda}  \longrightarrow  A_{\mu\nu\lambda} - \frac{1}{m}\, \big (\partial_\mu \phi_{\nu\lambda} + \partial_\nu \phi_{\lambda\mu}
+ \partial_\lambda \phi_{\mu\nu} \big )  
\equiv  A_{\mu\nu\lambda} - \frac{1}{m}\, \Sigma_{\mu\nu\lambda},
\end{eqnarray}
converts the second-class constraints of the
original Lagrangian density (23) into the first-class variety
due to the presence  of the St$\ddot u$ckelberg antisymmetric (i.e. $\phi_{\mu\nu} = -\, \phi_{\nu\mu}$) tensor
field  $\phi_{\mu\nu}$. The {\it latter} constraints are responsible for the  continuous, local
and infinitesimal gauge transformations: $\delta_g A_{\mu\nu\lambda} = 
(\partial_\mu \Lambda_{\nu\lambda} + \partial_\nu \Lambda_{\lambda\mu}
+ \partial_\lambda \Lambda_{\mu\nu}), \; \delta_g \phi_{\mu\nu} = m \Lambda_{\mu\nu} - 
(\partial_\mu\, \Lambda_\nu - \partial_\nu\, \Lambda_\mu), \; \delta_g \Sigma_{\mu\nu\lambda} = m \, 
(\partial_\mu \Lambda_{\nu\lambda} + \partial_\nu \Lambda_{\lambda\mu} + \partial_\lambda \Lambda_{\mu\nu}) , 
\; \delta_g H_{\mu\nu\lambda\zeta} = 0$ where
$\Lambda_{\mu\nu} $ and $\Lambda_\mu $ are the antisymmetric (i.e. $\Lambda_{\mu\nu} = -\, \Lambda_{\nu\mu} $)
tensor and vector infinitesimal gauge symmetry transformation parameters, respectively.
The following St$\ddot u$ckelberg-modified Lagrangian density (${\cal L}_{(A)}^{(S)}$),
obtained from  (23) by using (24), namely;
\begin{eqnarray}
{\cal L}_{(A)}^{(S)} = -\,\frac{1}{48}\, H^{\mu\nu\lambda\zeta}\, H_{\mu\nu\lambda\zeta} + \frac{m^2}{12}\, A^{\mu\nu\lambda}\, A_{\mu\nu\lambda}
- \frac{m}{6}\, A^{\mu\nu\lambda}\, \Sigma_{\mu\nu\lambda} + \frac{1}{12}\, \Sigma^{\mu\nu\lambda}\, \Sigma_{\mu\nu\lambda},
\end{eqnarray}
respects (i.e. $\delta_g {\cal L}_{(A)}^{(S)} = 0 $) the {\it above} infinitesimal gauge symmetry transformations in any 
arbitrary D-dimension of spacetime where one of the decisive features 
is the observation that the field-strength tensor
$H_{\mu\nu\lambda\zeta}$ does {\it not} change (i.e. $H_{\mu\nu\lambda\zeta} \rightarrow H_{\mu\nu\lambda\zeta} $)  under (24)

In the specific four (3 + 1)-dimensions of spacetime, the standard St$\ddot u$ckelberg-technique of 
replacement (24) can be modified on the basis of solid mathematical grounds as\footnote{Our four (3 + 1)-dimensional (4D)
Minkowskian flat spacetime is endowed with the totally antisymmetric Levi-Civita tensor $\varepsilon_{\mu\nu\lambda\zeta}$
where (i) the Greek indices: $\mu, \nu, \lambda, \zeta...= 0, 1, 2, 3$, and (ii) the convention chosen by us is
such that: $\varepsilon_{0123} = +\, 1 = -\,\varepsilon^{0123}$ along with the standard relationships:
$\varepsilon_{\mu\nu\lambda\zeta}\, \varepsilon^{\mu\nu\lambda\zeta} = -\, 4!, \; 
\varepsilon_{\mu\nu\lambda\zeta} \,\varepsilon^{\mu\nu\lambda\rho} = -\, 3!\, \delta^\rho_\zeta, \;
\varepsilon_{\mu\nu\lambda\zeta} \, \varepsilon^{\mu\nu\rho\sigma} = -\, 2!\, (\delta^\rho_\lambda \, \delta^\sigma_\zeta
- \delta^\rho_\zeta\, \delta^\sigma_\lambda)$, etc. The other notations 
as well as conventions, in the case of our {\it present} 4D theory, are similar to our {\it last} two sections.}
\begin{eqnarray}
A_{\mu\nu\lambda}  \longrightarrow    A_{\mu\nu\lambda} - \frac{1}{m}\, \Sigma_{\mu\nu\lambda}
- \frac{1}{m}\, \varepsilon_{\mu\nu\lambda\zeta} \, \partial^\zeta  \tilde \phi,
\end{eqnarray}
where the field $\tilde \phi$ is the pseudo-scalar field which ensures the existence of the parity symmetry in our theory
(unlike the theory of weak interaction where the parity symmetry is violated). The dimension four is very {\it special} for our
modified massive 4D Abelian 3-form theory because (i) the field-strength tensor    
$H_{\mu\nu\lambda\zeta}$ of our theory has only a {\it single} independent  
component which is nothing but $H_{0123}$, (ii) the kinetic term\footnote{The patterns of the {\it final} forms 
of the kinetic terms for
the modified 2D Proca theory (i.e. $-\, \frac{1}{4}\, F^{\mu\nu}\, F_{\mu\nu} = \frac{1}{2} \, F_{01}\,F_{01} $),
 modified 3D massive Abelian 2-form theory (i.e. $+\, \frac{1}{12}\, H^{\mu\nu\lambda}\, H_{\mu\nu\lambda} 
= \frac{1}{2} \, H_{012}\,H_{012} $)   and modified 4D massive Abelian 3-form theory
(i.e. $-\, \frac{1}{48}\, H^{\mu\nu\lambda\zeta}\, H_{\mu\nu\lambda\zeta} = \frac{1}{2} \, H_{0123}\, H_{0123}$)  are exactly the {\it same}. }
of our modified massive 4D Abelian 3-form 
theory becomes: $-\, \frac{1}{48}\, H^{\mu\nu\lambda\zeta}\, H_{\mu\nu\lambda\zeta} = \frac{1}{2} \, H_{0123}\, H_{0123}$, 
and (iii) the specific component
(i.e. $H_{0123}$)  does {\it not} remain invariant under the  modified
version  of the St$\ddot u$ckelberg-technique of replacement (26) because we observe the following, namely;
\begin{eqnarray}
H_{0123} \longrightarrow H_{0123} + \frac{1}{m}\; (\partial_0^2 - \partial_1^2 - \partial_2^2- \partial_3^2)\, \tilde \phi \;\equiv\; 
 H_{0123} + \frac{1}{m}\; \Box\, \tilde \phi,
\end{eqnarray}
which is {\it not} the case with the standard St$\ddot u$ckelberg-technique of replacement (24) where we have {\it already} noted that:
$H_{\mu\nu\lambda\zeta} \rightarrow H_{\mu\nu\lambda\zeta} $. Ultimately, the substitution of (26) into (23) leads to the derivation of 
the Lagrangian density (${\cal L}_{(A)}^{(MS)}$) as follows
\begin{eqnarray}
{\cal L}_{(A)}^{(MS)} &=& \frac{1}{2}\, (H_{0123} + \frac{1}{m}\; \Box\, \tilde \phi)^2 + \frac{m^2}{12}\, A^{\mu\nu\lambda}\, A_{\mu\nu\lambda}
- \frac{m}{6}\,  A^{\mu\nu\lambda}\, \bigl [\Sigma_{\mu\nu\lambda} + \varepsilon_{\mu\nu\lambda\zeta}\, \partial^\zeta \tilde \phi \bigr ]
\nonumber\\
&+& \frac{1}{12}\, \Sigma^{\mu\nu\lambda}\, \Sigma_{\mu\nu\lambda} - \frac{1}{2}\, \partial_\mu \tilde \phi\, \partial^\mu \tilde \phi, 
\end{eqnarray}
modulo a total spacetime derivative: 
$ \partial_\mu  \, [ \frac{1}{2}\,\varepsilon^{\mu\nu\lambda\zeta}\, \phi_{\nu\lambda} \, \partial_\zeta \, \tilde \phi ]$ which
does {\it not} play any significant role in the description of dynamics in our theory. The superscript $(MS)$ on the above
Lagrangian density denotes that we have taken into account the 
{\it modified} St$\ddot u$ckelberg-technique of replacement (26) in {\it its} derivation.
A few noteworthy points, at this stage, are as follows.
First, the kinetic term for the pseudo-scalar field turns-up with a {\it negative} sign. Second, we encounter the
higher derivative terms in the full expansion of the kinetic term (i.e. the first
term in the above Lagrangian density) for the gauge field. We can get rid of the {\it latter} problem by using
the on-shell condition (i.e. $\Box \, \tilde \phi = - m^2\, \tilde \phi$) for the pseudo-scalar field $\tilde \phi $
(cf. Sec. 2 for detail discussions). The ensuing
well-defined 4D Lagrangian density for our theory (without any kinds of higher order derivative terms)  is as follows:
\begin{eqnarray}
{\cal L}_{(A)}^{(MS)} &=& \frac{1}{2}\, (H_{0123} - m\;  \tilde \phi)^2 + \frac{m^2}{12}\, A^{\mu\nu\lambda}\, A_{\mu\nu\lambda}
- \frac{m}{6}\,  A^{\mu\nu\lambda}\, \bigl [\,\Sigma_{\mu\nu\lambda} + \varepsilon_{\mu\nu\lambda\zeta}\, \partial^\zeta \tilde \phi \; \bigr ]
\nonumber\\
&+& \frac{1}{12}\, \Sigma^{\mu\nu\lambda}\, \Sigma_{\mu\nu\lambda} - \frac{1}{2}\, \partial_\mu \tilde \phi\, \partial^\mu \tilde \phi. 
\end{eqnarray} 
One of the key consequences (that emerges out from the above Lagrangian density) is the observation that we obtain the EL-EoM w.r.t.
the pseudo-scalar field $\tilde \phi $ as
\begin{eqnarray}
\Box \, \tilde \phi - \frac{m}{6}\, \varepsilon^{\mu\nu\lambda\zeta}\,\partial_\mu\, A_{\nu\lambda\zeta} 
= m\, H_{0123} - m^2 \, \tilde \phi \;\;\; \Rightarrow \;\;& \big (\Box + m^2 \big)\, \tilde \phi = 0,
\end{eqnarray}
where we have used the following useful and important relationship:
\begin{eqnarray}
\frac{1}{6}\, \varepsilon^{\mu\nu\lambda\zeta}\,\partial_\mu\, A_{\nu\lambda\zeta} = 
\frac{1}{4!}\, \varepsilon^{\mu\nu\lambda\zeta}\,H_{\mu\nu\lambda\zeta} \equiv - \, H_{0123}.
\end{eqnarray}
The derivation of the EL-EoM (30) from the Lagrangian density (29) demonstrates that the 
on-shell condition (i.e. $\Box \, \tilde \phi = - m^2\, \tilde \phi$) that we have used 
for the derivation of the Lagrangian density (29) from (28), is correct. Finally, we note that the action integral
corresponding to the Lagrangian density (28) {\it still} 
respects the gauge symmetry transformations: $\delta_g A_{\mu\nu\lambda} = 
(\partial_\mu \Lambda_{\nu\lambda} + \partial_\nu \Lambda_{\lambda\mu}
+ \partial_\lambda \Lambda_{\mu\nu}), \; \delta_g \phi_{\mu\nu} = m \Lambda_{\mu\nu} - 
(\partial_\mu\, \Lambda_\nu - \partial_\nu\, \Lambda_\mu), \; \delta_g \Sigma_{\mu\nu\lambda} = m \, 
(\partial_\mu \Lambda_{\nu\lambda} + \partial_\nu \Lambda_{\lambda\mu} + \partial_\lambda \Lambda_{\mu\nu}) , 
\; \delta_g H_{\mu\nu\lambda\zeta} = 0, \; \delta_g \tilde \phi = 0$ (where the pseudo-scalar field does {\it not} transform at all).

For the quantization of our theory, we have to incorporate the proper gauge-fixing terms for (i) the gauge field ($A_{\mu\nu\lambda}$),
(ii) the St$\ddot u$ckelberg antisymmetric tensor field ($\phi_{\mu\nu}$), and (iii) the vector field ($\phi_{\mu}$). The {\it latter}
field exists in the theory  due to the reducibility properties of our massive Abelian 3-form gauge theory. The final form of the
gauge-fixed Lagrangian density (${\cal L}_{gf}^{(A)} $) for our theory is as  follows\footnote{In the Lagrangian densities (23) and (32), we have chosen the 
overall numerical and sign
factors which are different from our earlier work on the BRST and superfield approach to the D-dimensional modified
massive Abelian 3-form theory [29]. 
However, these changes have {\it not} affected the key EL-EoMs and the Curci-Ferrari type restrictions
that are present [29] in the bosonic sector of our BRST-quantized theory. } (see, e.g. [29] for details)
\begin{eqnarray}
{\cal L}_{gf}^{(A)} &=& \frac{1}{2}\, \Bigl [\partial^\lambda A_{\lambda\mu\nu}  + \frac{1}{2}\, (\partial_\mu \phi_\nu 
- \partial_\nu \phi_\mu) - \frac{m}{2}\, \phi_{\mu\nu} \Bigr ]^2 + \frac{1}{2}\, \Bigl [\frac{1}{2}\, 
\big (m\, \phi_\mu - \partial_\mu \phi \big ) 
- \partial^\nu \phi_{\nu\mu} \Bigr ]^2 
\nonumber\\ &-& \frac{1}{8}\, \Big (\partial \cdot \phi + m \, \phi \Big)^2,
\end{eqnarray}
where (i) the superscript $(A)$ on the above Lagrangian density denotes that we are focusing here on the modified
massive Abelian 3-form field $A_{\mu\nu\lambda}$, and (ii) the scalar field $\phi$ appears in the theory
due to the reducibility properties of the massive Abelian 3-form gauge field. Three key prosperities of the scalar field
(that are relevant to us at this stage)  are (i) it carries the kinetic term which has {\it positive} sign, (ii) it satisfies the 
usual KG equation of motion [i.e. $(\Box + m^2)\, \phi = 0$] which can be derived from the above gauge-fixing portion of the
total gauge-fixed Lagrangian density (i.e. ${\cal L}_{(A)}^{(MS)} + {\cal L}_{gf}^{(A)} $), and
(iii) it appears in our theory {\it only} in the gauge-fixing terms. Thus, it is clear that the scalar field $\phi$ of our
theory corresponds to the {\it real} massive relativistic elementary particle in contrast to the ``exotic'' pseudo-scalar field that 
corresponds to the ``phantom'' field and/or  a possible candidate for dark matter. The {\it latter}, to be precise,
is nothing but a pseudo-scalar elementary particle.

We conclude this section with the following remarks. First of all, properly gauge-fixed Lagrangian densities for {\it all}
the modified massive Abelian $p$-form (i.e. $p = 1, 2, 3  $) gauge theories contain the scalar as well as
the  pseudo-scalar fields. Whereas the
former field corresponds to a ``real'' massive relativistic elementary particle, the latter turns out to be an
``exotic'' version of it. Hence, the {\it latter} is a possible  candidate for dark matter (see, e.g. [17,18,30-34]). Second, we note that, in the cases of 
modified massive Abelian {\it higher} $p$-form (i.e. $p = 2, 3 $) gauge theories, the sources of the EL-EoMs for the scalar and PS
fields are {\it different} portions of the total gauge-fixed Lagrangian densities. The purely gauge-fixing portion leads to the
derivation of the KG equation for the scalar field and the {\it modified} portions 
[cf. Eqs. (20),(29)] of the total Lagrangian densities (where the MSF has been used) are responsible for the derivation of the KG equation
for the PS field. Finally, the distinguishing feature between the scalar and pseudo-scalar fields
of our theories is nothing but the positive and negative signs, associated 
with their kinetic terms, respectively.


\section{Conclusions}

In our present investigation, we have concentrated on the modifications of the {\it standard} St$\ddot u$ckelberg-technique of replacement
for the {\it massive} Abelian $p$-form ($p = 1, 2, 3$) fields in 
the contexts of the $D = p + 1$ dimensional (i.e. 2D, 3D, 4D) massive Abelian gauge
theories and established the existence of a {\it single} PS field that carries (i) the negative kinetic term, and (ii) a well defined
rest mass. The {\it latter} is equal to the rest mass of the 
{\it specific} massive  Abelian  $p$-form gauge field which is described by 
the {\it appropriate}  Lagrangian density
(that is defined on the $ D = p + 1$-dimensional flat Minkowskian  spacetime manifold). Thus, it is obvious that the 
origin of the PS field lies in the {\it modifications} of the standard St$\ddot u$ckelberg-technique of replacement
in the contexts of the massive $(p + 1)$-dimensional Abelian $p$-form gauge theories.
 This is one of the key observations of our present endeavor.

On the other hand, the existence of the pure scalar field owes its origin to (i) the proper gauge-fixing terms in the
contexts of the modified massive Abelian {\it higher} $p$-form ($p = 2, 3 $) gauge theories, and (ii) the {\it standard} St$\ddot u$ckelberg-technique 
of replacement for the massive Abelian 1-form gauge field [cf. Eqs. (2),(3)].
In our present endeavor, we have been
able to demonstrate that (i) the  massive pure scalar field appears in our  theories with a positive kinetic term, and
(ii) the PS field carries the negative 
sign with its kinetic term. However, {\it both} these fields obey the standard KG equation which
is derived from the properly gauge-fixed Lagrangian densities of the {\it above} modified 
massive ($p + 1$)-dimensional (2D,3D,4D) Abelian $p$-form ($p = 1,2,3$) gauge theories
(where MSF has been used). Thus, we have been able to establish that the PS field
is an ``exotic'' field (due to its {\it negative} kinetic term) which
contributes {\it only} to the gravitational interaction as it carries a properly well-defined mass. Such kinds of fields,
with the {\it negative} kinetic terms and the well-defined rest masses, are a set of possible candidates for (i) the dark matter [17,18,30-34], and
(ii) the ``phantom'' fields (see, e.g. [19-21]) as they are {\it different} from the
{\it usual} massive pure scalar field (that is described by the Lagrangian density  with the {\it positive} kinetic term and a rest mass).

In our Secs. 3 and 4, we have concentrated on the 3D modified massive Abelian 2-form and  4D modified massive Abelian 3-form
theories, respectively, because the higher $p$-form {\it basic} fields appear in the 
quantum excitations of the (super)string theories 
(see, e.g, [35-39]) which
are, at present, the forefront areas of research activities  
in the realm of theoretical high energy physics. The modern developments
in {\it these} theories have led to many new developments in the domain of research activities in pure mathematics. 
As far as the realm of quantum field theories are concerned, there is convergence and confluence of ideas from the pure
mathematics and theoretical physics in the discussions of the topological field theories 
(see, e.g. [40,41] and references therein), supersymmetric Yang-Mills theories (see, e.g. [42] and references therein), etc. 
We have purposely chosen the {\it modified} massive Abelian $p$-form ($p = 1, 2, 3 $) gauge theories {\it together} in the specific 
$D = p + 1$ dimensions ($D = 2, 3, 4 $) of spacetime so that we could establish that a {\it single} ``exotic'' PS field is present {\it not} only
in the context of the 2D modified massive Abelian 1-form (i.e. $p = 1 $) theory [22,27,28] {\it but} its existence is inevitable in 
the contexts of the modified massive Abelian {\it higher} $p$-form (i.e. $p = 2, 3$) theories, too, in the specific 
$ D = p + 1$  dimensions (i.e. $D = 3, 4$) of spacetime. Hence, this ``exotic'' field 
(i.e. the PS field) appears to be very {\it fundamental}
as it exists in the contexts of {\it all} the above modified $(p + 1)$-dimensional 
(i.e. 2D, 3D, 4D) {\it modified} massive Abelian $p$-form (i.e. $p = 1, 2, 3$) gauge theories.

Our present endeavor is {\it decisively} different from the massive models of the Hodge theory (see, e.g. [12])
where a couple of ``exotic'' fields (see, e.g. footnote on page 2) exist with the negative kinetic terms and well-defined masses.
One of the highlights of our present investigation is the observation that {\it only} a single PS field
(corresponding to an elementary pseudo-scalar particle) is needed for the modified massive Abelian $p$-form gauge theories 
in the specific $D = p + 1$ dimensions of spacetime with (i) the negative kinetic term, and
(ii) a well-defined rest mass. In other words, it appears to us that the PS field is the most
fundamental field which corresponds to the dark matter [17,18,30-34]. It is also the most basic object that would
correspond to the ``phantom'' and/or the ``ghost'' field in the realm of the cosmological models of the Universe [19-21,43-46].

We speculate that the {\it massless} limit of this basic ``exotic'' PS field (i.e. a possible dark energy candidate with
{\it only} a negative kinetic term) might a possible way to explain the current 
observational evidence of the accelerated cosmic expansion of the Universe (see, e.g. [47-51] for details). To corroborate our statement, 
we would like to lay emphasis on the {\it opposite} signs of the kinetic terms for the (pseudo-)scalar fields in our theory. The {\it massless} 
pseudo-scalar field, with the negative kinetic term, leads to the existence of the negative pressure in our theory which is one of
the key characteristic features of dark energy. Hence, the ``exotic'' PS field is one of the possible candidates for dark energy which is supposed to
be the reason behind the current experimental observation of the accelerated cosmic expansion of the Universe. On the contrary, the {\it positive}
kinetic term, associated with the pure scalar field, indicates that it is  a {\it normal}} massless field which does {\it not} generate negative pressure
in our theory. Thus, we conclude that the massless``exotic'' PS field (with the negative kinetic term) is one of the possible candidates for dark
energy which is responsible for the observation of the accelerated cosmic expansion of the Universe.

In our future endeavor, we plan to devote time on the proof of our 3D modified massive Abelian 2-form theory and 
4D modified massive Abelian 3-form theory to be the tractable 
field-theoretic examples for Hodge theory within the framework of BRST formalism
as we have already done in the context of the 2D {\it modified} Proca theory (see. e.g. [27,28,22] and references therein). It is gratifying to state that
(i) in  very recent work [52], we have been able to apply the BRST approach to discuss various aspects of 
the 3D {\it modified} massive Abelian 2-form theory with a {\it single} PS field, and (ii) in another recent paper [53], we have 
been able to prove a 3D {\it combined} system of the free Abelian 1-form and 2-form gauge theories to be a tractable field-theoretic
example for Hodge theory. This {\it specific} system has been shown to be the limiting case of our earlier work [52] with a particular identification and
a couple of stringent conditions. In yet another very recent  work [54], we have derived the algebraic structures  of the {\it appropriate} conserved charges
corresponding to the continuous symmetries of our earlier work [53] and established that there is two-to-one mapping between the {\it above}
charges and the cohomological operators of differential geometry. We have used the 't Hooft gauge-fixing term  (see, e.g. [24] for details) in the context
of {\it modified} 2D Proca theory where there is a U(1) gauge invariance. In an interesting work [55], the U(1) gauge symmetry, chiral symmetry and
existence of mass have been discussed. It will be nice future endeavor  to adopt this kind of approach in the contexts of the 
Abelian theories where there is existence of the U(1) gauge symmetry transformation. \\

\vskip 0.3cm

\noindent
{\bf Acknowledgments}\\

\vskip 0.3cm

\noindent
A key part of our present investigation was carried out at the School of Physics (SoP), University of Hyderabad (UoH)
when one of us (RPM) was invited to visit the SoP, UoH. The travel support and warm local hospitality, under the IoE-UoH- IPDF (EH) 
scheme of UoH, are thankfully acknowledged by him. Both the authors are grateful to the Reviewers for their very useful remarks.\\

\vskip 0.3cm

\noindent
{\bf  Data Availability Statement}\\

\vskip 0.3cm

\noindent
Data sharing not applicable to this article as no datasets were generated or analysed during the current study.\\

\vskip 0.3cm

\noindent
{\bf Conflicts of Interest} \\

\vskip 0.3cm

\noindent
The authors declare that there are no conflicts of interest. \\





\vskip 0.3cm

\noindent
{\bf Authors Confirmation}\\

\noindent
Both the authors agree to be accountable for the research presented.\\

\vskip 0.3cm

\noindent
{\bf Declaration}\\

\noindent
Both the authors declare that this manuscript is available as a preprint on the arXiv whose link is: https://arxiv.org/pdf/2402.11598.


\begin{thebibliography}{99}
\bibitem{RPM1}   T. Eguchi, P. B. Gilkey,  A. Hanson, Gravitation, gauge theories and differential geometry. Phys. Rep. 66, 213 (1980)
\bibitem{RPM2}   S. Mukhi, N. Mukunda, {\it Introduction to Topology, Differential Geometry and Group Theory for Physicists} 
                 (Wiley Eastern Private Limited, New Delhi, 1990)
\bibitem{Hari3}  K. Nishijima, The Casimir operator in the representations of BRS algebra. Prog. Theor. Phys. 80, 897 (1988)
\bibitem{Hari4}  J.W. van Holten, The BRST complex and the cohomology of compact lie algebras. Phys. Rev. Lett. 64, 2863 (1990)
\bibitem{Hari5}  M. G${\ddot o}$ckeler, T. Sch${\ddot u}$cker, 
                 {\it Differential geometry, gauge theories and gravity} (Cambridge University Press, Cambridge, 1987)               
\bibitem{RPM3}   C. Becchi, A. Rouet, R. Stora, The Abelian Higgs-Kibble model: unitarity of the S-operator. Phys. Lett. B 52, 344 (1974)
\bibitem{RPM4}   C. Becchi, A. Rouet, R. Stora, Renormalization of the Abelian Higgs-Kibble model. Comm. Math. Phys. 42, 127 (1975)
\bibitem{RPM5}   C. Becchi, A. Rouet, R. Stora, Renormalization of gauge theories.\\ Ann. Phys. (N. Y.) 98, 287 (1976)
\bibitem{RPM6}   I. V. Tyutin, Gauge invariance in field theory and statistical physics in operator formalism, 
                 in Lebedev Institute Preprint, Report Number: FIAN-39
                 (1975) (unpublished), arXiv:0812.0580 [hep-th]
\bibitem{RPM7}   Saurabh Gupta, R. P. Malik, Rigid rotor as a toy model for Hodge theory.\\ Eur. Phys. J. C 68, 325 (2010)
\bibitem{RPM8}   Shri Krishna, R. P. Malik, A quantum mechanical example for Hodge theory.\\
                 Ann. Phys. 464, 169657 (2024)
\bibitem{RPM9}   S. Krishna, R. Kumar, R. P. Malik,  A massive field-theoretic model for Hodge theory. Ann. Phys. 414, 168087 (2020)
\bibitem {RPM10} E. Harikumar, R. P. Malik, M. Sivakumar, Hodge decomposition theorem for Abelian two-form gauge theory.
                 J. Phys. A: Math. Gen. 33,   7149 (2000)
\bibitem{RPM11}  R. Kumar, S. Krishna, A. Shukla, R. P. Malik, Abelian p-form (p = 1, 2, 3) gauge theories as the field theoretic models for 
                 the Hodge theory.\\ Int. J. Mod. Phys. A 29, 1450135 (2014)
\bibitem{RPM12}  R. Kumar, R. P. Malik, Novel discrete symmetries in the general  N = 2 supersymmetric quantum mechanical model.
                 Eur. Phys. J. C 73, 2514  (2013)  
\bibitem{HARI1}  S. Krishna, R. P. Malik, A free supersymmetric system: novel symmetries.\\
                 Euro. Phys. Lett. 109, 31001 (2015)  
\bibitem{HARI2}  V. M. Zhuravlev, D. A. Kornilov, E. P. Savelova, 
                 The scalar fields with negative kinetic energy, dark matter and dark energy. Gen. Relat. Gravity 36, 1736 (2004)
\bibitem{HARI3}  Y. Aharonov, S. Popescu, D. Rohrlich, L. Vaidman, Measurements, errors, and negative kinetic energy. Phys. Rev. A 48, 4084 (1993)
\bibitem{HARI4}  P. J. Steinhardt, N. Turok, A cyclic model of the Universe. Science 296, 1436 (2002)
\bibitem{HARI5}  Y. F. Cai, A. Marcian, D.-G. Wang, E. Wilson-Ewing, Bouncing cosmologies with dark matter and dark energy. Universe 3, 1 (2017)
\bibitem{HARI6}   K. Koyama, Ghost in self-accelerating Universe. Class. Quant. Grav. 24, R231 (2007)
\bibitem{HARI19}  A. K. Rao, R. P. Malik, Modified Proca theory in arbitrary and two dimensions.\\ Euro. Phys. Lett. 135, 21001 (2021)
\bibitem{HARI20}  A. K. Rao, R. P. Malik, Modified St$\ddot u$ckelberg formalism: free massive Abelian 2-form theory in 4D.
                  Universe 9, 191 (2023)  
\bibitem{HAR21}   H. Ruegg, M. Ruiz-Altab, The stueckelberge field. Int. J. Mod. Phys. A 19, 3265 (2004)               
\bibitem{HARI22}  P. A. M. Dirac, {\it Lectures on Quantum Mechanics}, Belfer Graduate
                  School of Science (Yeshiva University Press, New York, 1964)                  
\bibitem{HARI23}   K. Sundermeyer, {\it Constrained Dynamics: Lectures Notes in Physics}, Vol. 169 \\
                  (Springer-Verlag, Berlin, 1982) 
\bibitem{RPM24}   B. Chauhan, S. Kumar, A. Tripathi, R. P. Malik, \\Modified 2D Proca theory: revisited under BRST and (anti-)chiral 
                  superfield formalisms. Adv. High Energy Phys. 2020,  3495168 (2020)
\bibitem{RPM25}   T. Bhanja, D. Shukla, R. P. Malik, Novel symmetries in the modified version of two dimensional Proca theory.
                  Eur. Phys. J. C 73   2535 (2013)            
\bibitem{HARI26}  A. K. Rao, R. P. Malik, 
                  Nilpotent symmetries of a modified massive Abelian 3-form theory: augmented superfield approach. Nucl. Phys. B 983, 115926 (2022)
\bibitem{HARI100} R. R. Caldwell, A phantom menace? Cosmological consequences of a dark energy component with super-negative equation of state.
                  Phys. Lett. B 545, 23 (2002)
\bibitem{HARI101} W. G. Gibbons, Phantom matter and the cosmological constant,\\ arXiv: hep-th/0302199
\bibitem{HARI102} S. M. Carroll, M. Hoffman, M. Trodden, Can the dark energy equation-of-state parameter will be less than -1?. 
                  Phys. Rev. D 68, 023509 (2003)   
\bibitem{HARI103} Mingzhe Li, Bo Feng, Xinmin Zhang, A single scalar field model of dark energy with equation of state crossing -1.
                  JCAP 0512, 002 (2005) 
\bibitem{HARI104}  R. R. Caldwell, Eric V. Linder, The limits of quintessence. \\ Phys. Rev. Lett. 95, 141301 (2005)         
\bibitem{HARI27}  M. B. Green, J. H. Schwarz, E. Witten, {\it Superstring Theory}, Vols. 1 and 2,\\
                  (Cambridge University Press, Cambridge, 1987)
\bibitem{RPM28}   J. Polchinski, {\it String Theory}, Vols. 1 and 2,\\
                  (Cambridge University Press, Cambridge, 1998)   
\bibitem{RPM29}   D. Lust, S. Theisen, {\it Lectures in String Theory} (Springer-Verlag, New York, 1989)
\bibitem{RPM30}    K. Becker, M. Becker, J.H. Schwarz, {\it String Theory and M-Theory} (Cambridge University Press, Cambridge, 2007)
\bibitem{RPM31}   D. Rickles, {\it A Brief History of String Theory From Dual Models to M-Theory} \\(Springer-Verlag, Germany, 2014)
\bibitem{RPM32}   E. Witten, Supersymmetric quantum mechanics on the lattice: I. Loop formulation. Nucl. Phys. B 202, 253 (1982)
\bibitem{RPM33}   A.S. Schwarz, On quantum fluctuations of instantons. Lett. Math. Phys. 2, 217 (1978)
\bibitem{RPM34}   Eric D'Hoker, D. H. Phong, {\it Lectures on Supersymmetric Yang-Mills Theory and Integrable Systems}, arXiv: 9912271 [hep-th]
\bibitem{RPM35}   M. Novello, S. E. P. Bergliaffa, Bouncing cosmologies. Phys. Rep. 463, 127 (2008)
\bibitem{RPM36}   J. L. Lehners, Ekpyrotic and cyclic cosmology. Phys. Rep. 465, 223 (2008)
\bibitem{RPM37}   Y. F. Cai, D. A. Easson, R. Brandenberger, Towards a nonsingular bouncing cosmology. JCAP 1208, 020 (2012)
\bibitem{RPM38}   S. Alexander, S. Cormack, M. Gleiser, A Cyclic universe approach to fine tuning.\\ Phys. Lett. B 757, 147 (2016)
 \bibitem{HARI35}  B. P. Schmidt,  Nobel Lecture: Accelerating expansion of the Universe through observations of distant supernovae.
                  Rev. Mod. Phys. 84, 1151 (2012)                     
\bibitem{HARI36}  P. Astier, R. Pain, Observational evidence of the accelerated expansion of the Universe,
                  arXiv: 1204.5493 [astro-ph.CO] 
\bibitem{HARI37}  Y. Gong, A. Wang, Energy conditions and current acceleration of the Universe.\\ Phys. Lett. B 652, 63 (2007) 
\bibitem{HARI105} F. D. Albareti, J. A. R. Cembranos, A. de la Cruz-Dombriz, Focusing of geodesic congruences in an accelerated expanding Universe.
                   JCAP 1212, 020 (2012) 
\bibitem{HARI106} M Tegmark, {\it et al}, Cosmological parameters from SDSS and WMAP.\\ Phys. Rev. D 69, 103501 (2004)
\bibitem{HARI38}  S. K. Panja, E. Harikumar, R. P. Malik, Modified massive Abelian 2-form theory with a single pseudo-scalar field: BRST approach, 
                  arXiv: 2405.15588 [hep-th]
\bibitem{HARI39}  A. K. Rao, R. P. Malik,  A 3D field-theoretic example for Hodge theory. \\Euro. Phys. Lett. (EPL)  147, 32001 (2024) 
\bibitem{HARI40}  Bhagya. R, Harsha Sreekumar,  E. Harikumar, R. P. Malik, Constraints, conserved charges and extended BRST algebra for a 3D 
                  field-theoretic example for Hodge theory, Nucl. Phys. B 1009, 116736 (2024)  
\bibitem{HARI222} E. Guendelman, Gauge invariance and mass without spontaneous symmetry breaking.   Phys. Rev. Lett. 43, 543 (1979) 
       
 \end{thebibliography}
\end{document}